# Nomenclature Ontology for Medical And Disease names (NOMAD): taxonomy of types and origins of disease names

Spiros Denaxas[1,2,4] *, Cai Ytsma[1], Giannos Louloudis[3], Jackie MacArthur[2,] and Harry Hemingway[1]

[1] *University College London, Institute of Health Informatics, NW1 2DA, London, United Kingdom*
[2] *British Heart Foundation Data Science Centre, NW1 2DE, London, United Kingdom*
[3] *University of Copenhagen, 1165, Copenhagen, Denmark*
[4] *Interdisciplinary Transformation University, A-4040, Linz, Austria*

**Abstract**
The nomenclature of human disease has developed organically over the past centuries using Greek, Latin, and Arabic terminology and reflects the idiosyncrasies of different eras of medical discovery. Despite evident heterogeneity in naming practices, no systematic framework exists for characterising these conventions across all diseases. In this paper, we describe the Nomenclature Ontology for Medical And Disease names (NOMAD), a meta-taxonomy that classifies disease names according to their naming conventions. We developed a two-level taxonomy comprising 9 top-level categories and 20 subcategories and applied it to 22,548 index entries from the ICD-10-CM 2026 Alphabetical Index in a scalable three-stage machine learning-driven classification pipeline. Classification was multi-label, reflecting the compositional nature of medical nomenclature. We classified 99.1% of terms with a mean of 2.12 labels per entry. Anatomical categories were the most prevalent (63.8% of entries), followed by Descriptive (48.4%) and Pathophysiological (40.2%), while Eponymous and Geographical labels were less common than their cultural prominence might suggest (9.7% and 1.9% respectively). Among all Eponymous diseases, we identified only 57 (2.6%) of diseases named after a female person. We manually reviewed a random sample of n=2,255 entries (10%) for accuracy and calculated a full agreement rate of 70% and partial agreement rate of 26% (macro-averaged Cohen's Kappa score 0.832). Naming convention profiles varied substantially across ICD-10-CM chapters, reflecting specialty-specific epistemological traditions: infectious disease chapters were dominated by etiological labels and showed the highest proportion of geographical region related labels, the circulatory chapter by anatomical and pathophysiological labels, and mental and behavioural disorders showed the highest prevalence of socio-behavioral labels. NOMAD provides an empirical foundation for understanding the structure of medical terminology with potential applications in clinical NLP, biomedical text mining, and terminology governance.

## 1. Introduction

"They certainly give very strange and newfangled names to diseases" Plato remarked, highlighting the confusing, archaic, or oddly specific terminology used in medicine since antiquity [1]. Historically, the nomenclature of human diseases developed organically rather than systematically, drawing heavily from the linguistic roots of early medical pioneers. Many foundational terms originated from ancient Greek and Latin, relying on anatomical prefixes combined with symptom-based suffixes (e.g., arthritis, derived from the Greek arthron for "joint" and -itis for "inflammation"). Later, Arabic physicians and scholars translated, synthesized, and expanded upon existing medical nomenclature [2].

As global medical research accelerated in the 19th and 20th centuries, naming conventions became increasingly idiosyncratic. Diseases were frequently designated using eponyms [3,4] to honor the physicians who first characterized them (e.g., Alzheimer's disease), the geographic locations of early identified outbreaks (e.g., Ebola virus disease, named after the Ebola River), or presumed animal vectors (e.g., Avian influenza). While historical naming practices provided convenient shorthand, they frequently resulted in unintended stigmatization, diplomatic friction, and economic disruptions for associated regions. As a result, the World Health Organization (WHO) issued guidance on the naming of infectious diseases and proposed avoiding geographical locations (e.g. Middle Eastern Respiratory Syndrome), eponyms (e.g. Chagas), cultural or occupational references (e.g. Legionnaire's disease), and animal or food species (e.g. Mad cow disease) [5]. These naming patterns and idiosyncrasies can still be traced in the terms used in modern medicine as part of statistical classification systems (International Classification of Disease) and medical ontologies (e.g. SNOMED).

Despite the long history and evident heterogeneity of disease naming practices, no systematic framework exists for classifying the conventions themselves i.e. for characterizing the linguistic and conceptual strategies that underpin medical nomenclature at scale. Namer and Baud [6] developed a morphosemantic system for automatically assigning semantic information to compound biomedical terms by decomposing them into morphologically meaningful units and relating these across languages. Hahn et al. [7] proposed subword segmentation methods for medical document retrieval, showing that decomposing morphologically complex terms into medically significant subwords improved retrieval performance over conventional string matching. These approaches treat term structure as a tool for information retrieval and cross-lingual mapping rather than as a window onto naming conventions themselves, but they establish that medical terminology has exploitable compositional structure.

Several disease ontologies provide structured representations of disease entities and their relationships but do not characterize the naming conventions that produced the terms. The Human Disease Ontology (DO) integrates disease classifications with etiological and phenotypic features [8]. The Mondo Disease Ontology harmonizes multiple disease resources including OMIM, Orphanet, and ICD into a unified ontological framework, with particular attention to cross-resource mapping [9]. The International Classification of Diseases Ontology (ICDO) [10] decomposes ICD terms into components including anatomical entities, process profiles, and etiological causes. While individual naming patterns (eponyms, anatomical terms, geographical references) have been discussed in isolation, often in the context of proposed reforms, the full landscape of how diseases acquire their names has not been mapped comprehensively. A structured representation of disease naming practices can be helpful in clinical NLP and biomedical text mining where understanding the structural patterns embedded in disease names could inform more robust entity recognition and normalization systems.

A term like "hereditary spherocytosis" encodes both an etiological signal and a morphological description; making such compositional structure explicit could improve the performance of systems that must resolve synonyms, parse unfamiliar terms, or link free text to standardized codes. From the perspective of terminology governance and harmonization, a systematic characterization of existing naming practices provides an empirical foundation for evaluating the coherence and consistency of coding systems such as ICD-10-CM, and for identifying where nomenclature may introduce ambiguity, cultural bias, or barriers to interoperability. In this paper, we present the Nomenclature Ontology for Medical and Disease names (NOMAD), a meta-taxonomy that systematically classifies disease names according to their naming conventions. Our objectives were:

1. to develop and validate a multi-level taxonomy of disease naming conventions encompassing eponymous, geographical, anatomical, etiological, pathophysiological, descriptive, socio-behavioral, and vernacular categories

2. to apply this taxonomy to the ICD-10-CM 2026 alphabetical index and characterize the distribution and co-occurrence of naming conventions across the full breadth of coded diseases

3. demonstrate the feasibility of a scalable, reproducible approach to nomenclature analysis that could be extended to other terminologies and coding systems.

## 2. Methods

### 2.1. Data source and pre-processing

We downloaded the 2026 version of the ICD-10-CM Alphabetical Index to Diseases and Nature of Injury XML document from the CDC National Center for Health Statistics (https://www.cdc.gov/nchs/icd/icd-10-cm/files.html). To reduce duplication and increase the interpretability of our analyses, we retained only level 0 and level 1 index entries for analyses. Furthermore, we excluded ICD-10-CM codes from chapters R (symptoms and signs), S/T (injury), V/Y (external causes), and Z (factors influencing health status and contact with health services) as their naming conventions are predominantly descriptive of clinical circumstances rather than disease entities.

### 2.2. Disease naming taxonomy

To our knowledge, no taxonomy of disease naming patterns exists and we therefore defined a two-level taxonomy of disease naming conventions comprising 9 top-level categories and 20 subcategories (**Table 1**). The top-level categories capture the primary conceptual strategy encoded in a disease name: Eponymous (named after a person), Geographical (named after a place), Anatomical (specifying a body site), Etiological (encoding a cause), Pathophysiological (encoding a disease process), Descriptive (characterizing a clinical feature or circumstance), Socio-Behavioral (encoding a social or psychological dimension), Vernacular/Obscure (colloquial, archaic, or non-standard terms), and Other/Unclassified. Four of these categories ( Eponymous, Anatomical, Etiological, and Descriptive) are further subdivided to capture finer-grained distinctions. For example, Etiological is subdivided into Infectious Agent, Substance-related, Injury-related, Genetic, Nutritional, Immune-mediated, Occupational, and Procedural, reflecting the diversity of causal attributions encoded in disease names.

**Table 1**: Taxonomy of disease naming conventions.

| Category | Subcategory | Naming convention |
|---|---|---|

| Eponymous | Physician/Scientist | Physician/scientist first described the condition |
|---|---|---|
| Eponymous | Other/Unknown | Patient, mythological figure, literary character, or person of uncertain identity |
| Geographical | | Geographic location e.g. country |
| Anatomical | Organ | Specifies an organ as the primary site affected |
| Anatomical | System | Specifies a body system e.g. respiratory |
| Anatomical | Tissue | Specifies a tissue type e.g. epithelial |
| Anatomical | Cellular | Specifies a cell type e.g. lymphoblastic |
| Etiological | Infectious Agent | Pathogen or infection term e.g. bacterial |
| Etiological | Substance-related | Drug, chemical, toxin, or alcohol as cause |
| Etiological | Injury-related | Trauma, radiation, or physical injury as cause |
| Etiological | Genetic | Hereditary, chromosomal, or gene-related cause |
| Etiological | Nutritional | Encodes a dietary deficiency or excess |
| Etiological | Immune-mediated | An immune mechanism e.g. autoimmune |
| Etiological | Occupational | A work-related cause or exposure |
| Etiological | Procedural | A medical or surgical procedure as cause |
| Pathophysiological | | Abnormal biological process e.g. fibrosis |
| Descriptive | Symptomatic | Symptom or clinical sign |
| Descriptive | Metaphorical | Uses a figurative or analogical term |
| Descriptive | Syndromic | Groups features under a collective label |
| Descriptive | Population | Affected population e.g. age group |
| Descriptive | Course/Severity | Temporal course or severity e.g. acute |
| Descriptive | Congenital | Terms encoding presence from birth |
| Socio-Behavioral | | Social circumstance, psychological state, or behavioral pattern |
| Vernacular/Obscure | | Colloquial/folk/archaic/foreign term |
| Other/Unclassified | | Name does not fit any other category (often abbreviations or highly generic terms) |

Classification is multi-label: a single disease name (as defined in the ICD-10-CM index) may be assigned to multiple categories when different components of the name independently encode different naming conventions (e.g., "Streptococcal pharyngitis" would be assigned both Etiological:Infectious Agent and Anatomical:Organ). A core principle of the taxonomy is that classification is based solely on what the name itself encodes, not on clinical knowledge of the underlying disease. For example, "Huntington's disease" is classified only as Eponymous, not as Etiological:Genetic, because the name does not contain any information with regards to the genetic aetiology of the disease.

### 2.3. Classification of disease names

To classify individual ICD-10-CM index terms, we implemented a three-stage approach that exploits the hierarchical structure of the index. In the first stage, we classified unique main terms (i.e. "Colitis", "Dermatosis", "Stroke"). In the second stage, we classified unique subterms independently (i.e. "joint", "mitral", "intestinal"). In the third and final stage, we created the final labels for each index entry by taking the union of labels assigned to its main term and subterm components. For example, "A52.16 Charcot's > joint" was assigned classes "Eponymous:Physician/Scientist" and "Anatomical:Organ" as they are the superset of individual classifications of the constituent parts of the ICD-10-CM description.

Classification was performed using Claude Sonnet (claude-sonnet-4-6) in batches of five index terms per batch using a custom Python v. 3.12 pipeline. The system prompt used is provided in **Appendix A**. To validate the quality of the LLM-based classification, we manually reviewed a random 10% sample of final label results and assigned an agreement score ("Agree", "Partial Agree", and "Disagree"). Based on these annotations, Cohen's κ was computed per taxonomy label by binarising each label independently (present/absent) and computing κ between LLM and annotator binary vectors; macro-averaged κ was then calculated across the 25 labels for which both raters assigned the label at least once. We used descriptive statistics to characterize the labelled dataset. We calculated and reported the mean and median number of labels assigned to each index term. We used heatmaps to visualize the distribution of labels and top-level classes across individual ICD-10-CM chapters.

### 2.4. Serialized data files

To facilitate reuse and integration with existing biomedical knowledge infrastructure, we serialized NOMAD in a format compatible with Web Ontology Language (OWL) 2 DL. The NOMAD class hierarchy directly reflects the two-level taxonomy described above. Each of the nine top-level categories was defined as a named OWL class, with subcategories represented as subclasses. All classes were assigned globally unique IRIs under a dedicated NOMAD namespace (https://w3id.org/nomad/ [registered after decision on paper]), persistent identifiers and human-readable annotations including rdfs:label, skos:definition, and skos:example drawn from the taxonomy definitions in **Table 1**. The ICD-10-CM index entries classified in this study were represented as OWL named individuals, with each individual linked to its corresponding NOMAD class(es) via rdf:type assertions, reflecting the multi-label nature of the classification. Each individual was additionally annotated with the ICD-10-CM code using a dedicated nomad:icd10cmCode and the source description string (via rdfs:label) to support downstream lookup and integration. All resources are released under an open source license at [https://github.com/spiros/nomad].

## 3. Results

We extracted 22,548 descriptions (index entries) from the ICD-10-CM index. These mapped to 6,396 unique ICD-10-CM codes, of which 18,095 (80.3%) contained a subterm (i.e., entries of the form "Main term > subterm"). We performed 14,193 API calls which represents a reduction of ~37% over performing 22,548 calls to classify entries one by one. A total of 4,453 (19.7%) entries consisted of main terms alone. In the first and second passes, unique main terms (n=5,765) and unique subterms (n=8,428) were classified independently. In the final stage, final labels for each index entry were derived by taking the union of labels assigned to its main term and subterm components. Our approach achieved near-complete coverage: 22,336 entries (99.1%) received at least one substantive category label. In total, 212 entries (0.9%) were assigned exclusively to "Other / Unclassified," typically abbreviations (e.g., "RAEB," "CADASIL," "PAT"), acronyms without expansion, or highly generic terms (e.g., "Anomaly, anomalous," "Problem") (**Table 2**).

**Table 2**: Classification of 22,548 ICD-10-CM index entries (2026 release) into a taxonomy of 25 naming convention categories across 9 top-level dimensions. N (%) reflects the number and proportion of entries assigned to each label. As entries may receive more than one label, column percentages sum to more than 100%. Representative ICD-10-CM terms and codes are shown for each category. Categories with no subcategory (e.g. Geographical, Pathophysiological) operate as atomic classes within the taxonomy.

| Label | Example disease [and description] | N entries (%) |
|---|---|---|
| Eponymous-Physician/Scientist | Alzheimer's disease (G30.9), named after the German physician Alois Alzheimer | 2,077 (9.2%) |
| Eponymous-Other/Unknown | Christmas disease (D67), named after Stephen Christmas, a 5-year-old boy who was the first person diagnosed with hemophilia B | 115 (0.5%) |
| Geographical | Ebola virus disease (A98.4), Ebola River in the Democratic Republic of the Congo | 435 (1.9%) |
| Anatomical:Organ | Nephritis (N05.9) | 8,561 (38.0%) |
| Anatomical:System | Anthrax > respiratory (A22.1) | 1,141 (5.1%) |
| Anatomical:Tissue | Dermatitis (L30.9), a common, non-contagious skin inflammation | 4,006 (17.8%) |
| Anatomical:Cellular | Lymphoma > lymphoblastic (C83.5-), cancer of immature lymphoblasts that arises in T cells | 671 (3.0%) |
| Etiological:Infectious Agent | Streptococcal pharyngitis (J02.0), after the Streptococcus bacteria | 2,885 (12.8%) |
| Etiological:Substance-related | Alcoholic dementia (F10.97), due to excessive alcohol consumption | 586 (2.6%) |
| Etiological:Injury-related | Radiation pneumonia (J70.0), due to excessive radiation exposure | 246 (1.1%) |
| Etiological:Genetic | Hereditary chorea (G10), Familial tremor (G25.0), Trisomy (Q92.9) | 288 (1.3%) |
| Etiological:Nutritional | Malnutrition (E46) | 371 (1.6%) |

| Etiological:Immune-mediated | Autoimmune purpura (D69.0), autoimmune mediated mechanism | 164 (0.7%) |
|---|---|---|
| Etiological:Occupational | Farmer's lung (J67.0), hypersensitivity pneumonitis caused by breathing in mold spores from damp, stored crops like hay | 188 (0.8%) |
| Etiological:Procedural | Graft-versus-host disease (D89.813), serious complication of allogeneic stem cell/ or bone marrow transplants | 320 (1.4%) |
| Pathophysiological | Osteonecrosis (M87.9) | 9,060 (40.2%) |
| Descriptive:Symptomatic | Fever (R50.9) | 5,574 (24.7%) |
| Descriptive:Metaphorical | Strawberry mark (Q82.5), a common benign vascular tumor in babies | 1,218 (5.4%) |
| Descriptive:Syndromic | HELLP syndrome (O14.2-) | 1,437 (6.4%) |
| Descriptive:Population | Neonatal acne (L70.4) | 921 (4.1%) |
| Descriptive:Course/Severity | Chronic gout (M1A.00) | 1,344 (6.0%) |
| Descriptive:Congenital | Congenital genu (Q74.1) | 428 (1.9%) |
| Socio-Behavioral | Xenophobia (F40.10) | 665 (2.9%) |
| Vernacular/Obscure | Croup (J05.0), Mumps (B26.9), Bejel (A65) | 1,879 (8.3%) |
| Other / Unclassified | CADASIL (I67.850), acronym for Cerebral Autosomal Dominant Arteriopathy with Subcortical Infarcts and Leukoencephalopathy | 212 (0.9%) |

To validate the classification, we created a random sample of 2,255 records (~10% of the entire dataset) and reviewed the results. Of the 2,255 reviewed entries, 70.0% received fully correct labels, 26.0% showed partial agreement, and 4.0% were fully incorrect. Near-perfect agreement ($\kappa > 0.95$) was observed for well-defined, unambiguous categories: Descriptive:Syndromic ($\kappa = 0.985$), Descriptive:Population ($\kappa = 0.964$), Eponymous:Physician/Scientist ($\kappa = 0.963$), Descriptive:Congenital ($\kappa = 0.960$), Etiological:Procedural ($\kappa = 0.953$), and Etiological:Infectious Agent ($\kappa = 0.946$). Lower agreement was noted for inherently ambiguous categories. Descriptive:Metaphorical ($\kappa = 0.650$) and Etiological:Nutritional ($\kappa = 0.652$) showed moderate agreement and the lowest agreement was for Vernacular/Obscure ($\kappa = 0.304$). In cases of partial agreement, the most common error was the LLM over-applying Vernacular/Obscure (added incorrectly in 101 cases) to terms that are standard medical vocabulary i.e. Cicatrix > prostate (N42.89), Catarrh, catarrhal > mouth (K12.1), and Furuncle > perineum (L02.225) were flagged as archaic or obscure when they are established clinical terms. Descriptive:Symptomatic was incorrectly added in 63 cases, most commonly applied to structural or pathological states rather than true symptoms (i.e., Microcolon (Q43.8), Levocardia (Q24.1), Dextrocardia (Q24.0), Polyp > colon (K63.5), and Absence > aorta (Q25.41) were flagged as symptomatic when they describe anatomical or pathological entities. A further source of partial error was confusion between Anatomical:Tissue and Anatomical:Organ (44 and 43 cases respectively), where the LLM assigned the wrong anatomical subcategory i.e. the femoral vein in Thrombophlebitis > femoral vein (I80.1) was classified as tissue rather than organ, while the nipple in Puerperal > thelitis (O91.02) was classified as tissue rather than organ. In fully discordant cases (4.0%), the dominant pattern was again over-application of Vernacular/Obscure (46 of 90 cases), where standard terms such as Furuncle > buttock (L02.32) and Urticaria > dermatographic (L50.3) were incorrectly flagged. Misclassification of -emia terms as Descriptive:Symptomatic rather than Pathophysiological was a recurring theme, affecting entries such as Hydroxyprolinemia (E72.59), Hypergastrinemia (E16.4), and Anemia > macrocytic (D53.9), where the LLM appeared to interpret abnormal blood-level terms as symptom descriptors rather than pathophysiological states. A further pattern involved incorrect assignment of Etiological:Nutritional to enzyme deficiencies i.e. Deficiency > mevalonate kinase (M04.1) — and unwarranted Geographical labelling based on etymological rather than explicit geographic reference, as in Thalassemia with other hemoglobinopathy (D56.8).

Classification was multi-label, reflecting the composite nature of medical nomenclature. A total of 47,793 label assignments were made across 22,548 entries. All 25 distinct subcategory labels from the taxonomy were represented in the output. The mean number of labels per entry was 2.12 (median: 2.0; range: 1–6). Most entries received exactly 2 labels (62.8%), followed by 3 labels (21.2%) and 1 label (13.8%). Only 2.2% of entries received 4 or more labels.

At the top level, Anatomical categories were the most prevalent, appearing in 63.8% of entries, followed by Descriptive (48.4%), Pathophysiological (40.2%), Etiological (22.4%), and Other/Unclassified (14.2%). Less common were Eponymous (9.7%), Vernacular/Obscure (8.3%), Socio-Behavioral (2.9%), and Geographical (1.9%). Across Eponymous diseases, we identified that 97.2% of entries were based on a male person, 2.6% were based on a female person (of which many with a male person included), and 0.1% entries where the gender could not be determined e.g. Hartnup Disease (E72.02) named after the Hartnup family, the first patients described with the condition. The extremely low proportion of female Eponyms is consistent with the established pattern of systematic under-recognition of female scientists ("Matilda Effect")[11]. At the subcategory level, the five most frequent labels were: Pathophysiological (40.2% of entries), Anatomical:Organ (38.0%), Descriptive:Symptomatic (24.7%), Anatomical:Tissue (17.8%), and Other/Unclassified (14.2%). Among etiological subcategories, Infectious Agent was dominant (12.8%), followed by Substance-related (2.6%), Nutritional (1.6%), Procedural (1.4%), Genetic (1.3%), Injury-related (1.1%), Occupational (0.8%), and Immune-mediated (0.7%). Within the Descriptive subcategories, Symptomatic was most common (24.7%), followed by Syndromic (6.4%), Course/Severity (6.0%), Metaphorical (5.4%), Population (4.1%), and Congenital (1.9%).

Analysis of label co-occurrence at the top level revealed that the most frequent combination was Anatomical + Pathophysiological, co-occurring in 32.7% (n=7,371) of entries. This reflects the common pattern of disease names that specify both a body site and a pathological process (e.g., "hepatic fibrosis," "renal infarction"). The next most common pairs were Anatomical + Descriptive (18.6%) (n=4,195) and Anatomical + Etiological (10.4%) (n=2,341). Notably, Descriptive + Eponymous co-occurred in 4.8% of entries, suggesting that many eponymous conditions retain descriptive modifiers in their index entries (e.g., "Friedreich's ataxia" combining an eponym with a symptom-based subterm).

The distribution of naming conventions varied substantially across clinical domains (represented as distinct ICD-10 chapters), consistent with the distinct traditions of nomenclature in different medical specialties (**Figure 1**). Infectious disease chapters (A00–B99; n=3,828) were dominated by Etiological labels (65–73% of entries), reflecting the convention of naming infections by their causative organisms. Circulatory system diseases (I00–I99; n=1,199) showed the highest prevalence of Anatomical labels (84% top labels), followed by Pathophysiological (61%), consistent with cardiovascular terminology that typically specifies both the affected vessel or chamber and the pathological process. Mental and behavioral disorders (F01–F99; n=1,395) were distinctive in showing the highest prevalence of Socio-Behavioral labels (44%) and the lowest reliance on anatomical naming. The Pregnancy chapter (O00–O9A; n=536) and Perinatal chapter (P00–P96; n=389) were overwhelmingly Descriptive (76.9% and 88.7% of entries respectively, reflecting multiple descriptive labels per entry), consistent with the temporal and population-specific framing of obstetric and neonatal terminology.

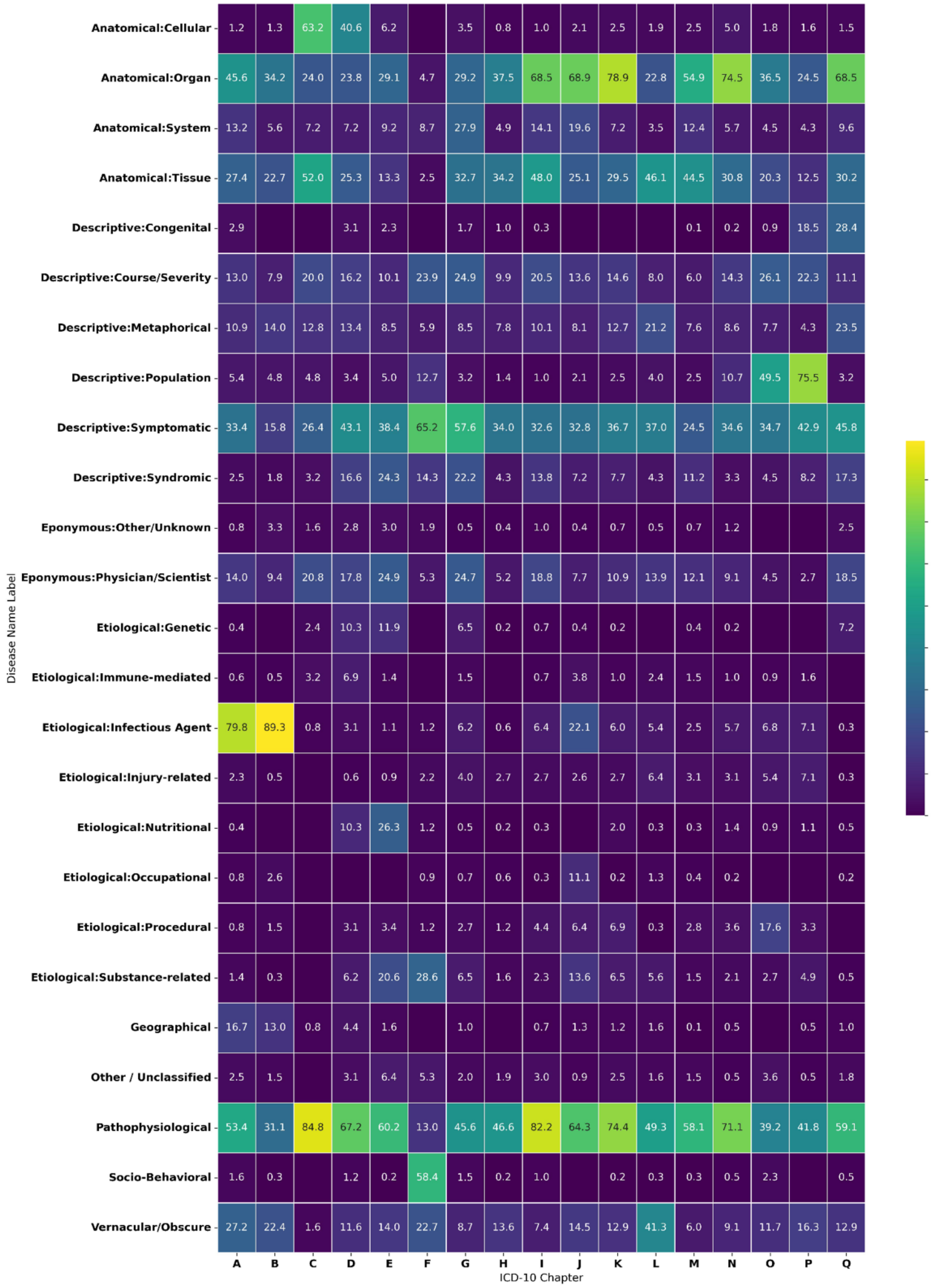

**Figure 1**: Heatmap representation of unique labels assigned to ICD-10-CM index terms stratified by ICD-10 chapter as defined by the first character of the ICD-10-CM code.

## 4. Discussion

Our classification reveals that ICD-10-CM is not a terminologically uniform system but an amalgamation of historically distinct naming logics, with anatomical and pathophysiological conventions providing the dominant structural scaffolding. The heavy usage of Anatomical labels (63.8% of all entries) reflects the extent to which body-site specificity has become the default organising principle of the classification, functioning less as a naming convention in the traditional sense and more as a baseline expectation onto which other conventions are layered (further evidenced by the fact that the three most common label co-occurrences all involve an Anatomical component).

Overall, the LLM-driven classifier demonstrated strong labelling accuracy across the taxonomy, with disagreements concentrated in subjective categories where definitional boundaries are less well defined. In 70% of the cases, the classifier assigned the correct labels. In entries where there was partial agreement, this was largely due to the classifier incorrectly assigning the label "Descriptive:Symptomatic" to several diseases (i.e. Q15.0 Megalocornea with glaucoma, Q43.8 Microcolon).

The prevalence of multi-label entries (mean 2.12 labels; 84.2% of entries receiving two or more) underscores that medical nomenclature is fundamentally compositional: most disease names do not belong to a single naming tradition but combine anatomical, etiological, descriptive, and pathophysiological signals within a single term, a complexity that is obscured when naming conventions are discussed in isolation. Conversely, the relative rarity of Eponymous (9.7%) and Geographical (1.9%) labels at the classification level is notable given the historical prominence of both in medical discourse. It suggests that, despite high-profile examples, eponymous and geographically-derived names constitute a small fraction of the overall nomenclature. This is broadly consistent with the direction of WHO naming reform, even if such names retain outsized cultural and clinical visibility.

Complex entries, i.e. entries receiving 5 or 6 labels, illustrate how medical naming conventions can layer multiple classificatory dimensions. For instance, "Pellagra-cerebellar-ataxia-renal aminoaciduria syndrome" (E72.02) received 6 labels: Vernacular/Obscure, Descriptive:Symptomatic, Anatomical:System, Anatomical:Organ, Pathophysiological, and Descriptive:Syndromic. These captured the name's use of an archaic term, multiple anatomical references, a pathological process, and syndromic framing. Similarly, "Stevens-Johnson disease or syndrome > toxic epidermal necrolysis overlap" (L51.3) received 5 labels: Eponymous, Syndromic, Substance-related, Anatomical, and Pathophysiological categories.

Our results show that naming conventions are not merely incidental features of disease terminology but reflect deep, specialty-specific epistemological traditions in how different branches of medicine have historically conceptualised and communicated disease. For example, infectious disease chapters (A and B) are overwhelmingly etiological ( Etiological:Infectious Agent labels appearing in 79.8% and 89.3% of entries respectively). This reflects a naming logic built around causative organisms that has been systematically reinforced since the germ theory era. We observed the opposite in the the circulatory chapter (I): 84% of entries carry an Anatomical label (almost entirely Anatomical:Organ (78.9%)) and 74.4% carry a Pathophysiological label, consistent with an approach that anchors disease identity in structure and process rather than cause.

The dermatology chapter (L) exhibits the highest prevalence of Vernacular/Obscure labels (41.3%) across the entire classification, substantially exceeding chapters with comparably ancient clinical traditions. This likely reflects the persistence of archaic, often Latinate or Greek-derived vernacular terms in dermatological nomenclature. Terms include psoriasis, urticaria, and ichthyosis that have resisted reformulation and reflect the historically descriptive, visually-anchored nature of skin disease diagnosis, where lesion morphology gave rise to figurative and colloquial language which was never systematised or modernized. Likewise, the near-invisibility of "Geographical" labels across most chapters (i.e. <2%) stands in sharp contrast to the heightened prevalence in chapters A and B (16.7% in A), underscoring that the WHO's long-standing concern about geographically-stigmatising disease names is concentrated almost exclusively in the infectious disease domain.

Our study has several limitations worth discussing. NOMAD was developed and applied exclusively to ICD-10-CM 2026, and its coverage is therefore restricted to naming conventions present in a single, English-language coding system. Whether the taxonomy and its distributional findings generalise to other terminological resources (e.g. SNOMED, ICD-11, Orphanet etc) remains to be established. We restricted analysis to level 0 (e.g. “Myocarditis”) and level 1 index entries (e.g. “Myocarditis > staphylococcal”). We excluded six ICD-10 chapters (R, S, T, V, Y, Z) whose naming conventions are dominated by clinical circumstance rather than disease identity; the full index may exhibit a somewhat different distributional profile. Classification relied on a single classifier without ensemble comparison or cross-model validation, and our error analysis identified a small percentage of incorrect classifications that future work should address through targeted prompt refinement or post-hoc rule-based correction. Finally, manual validation covered a 10% random sample; while this is proportionally substantial, rarer categories are represented by smaller absolute counts and kappa estimates for low-frequency labels such as Vernacular/Obscure should be interpreted with caution.

## 5. Conclusion

NOMAD provides the first systematic empirical characterisation of disease naming conventions at scale, revealing that medical nomenclature is fundamentally compositional, specialty-specific in its epistemological traditions, and far more anatomically anchored than cultural discourse around eponyms and geographical names might suggest.

## A. System prompt used for classification

You are an expert in medical terminology and disease nomenclature. Your task is to classify disease names according to their **naming conventions** — the linguistic and conceptual strategies used to coin each name.

**Core Principle**

Classify based ONLY on what the name itself encodes. Do NOT apply categories based on your clinical knowledge of the underlying disease. The question to ask for every label is: "Does the name itself contain a word or element that signals this category?"

- CORRECT: "Streptococcal pharyngitis" → `Etiological:Infectious Agent` (the name says "streptococcal") + `Anatomical:Organ` (pharynx)
- CORRECT: "Down syndrome" → `Eponymous:Physician/Scientist` only. Do NOT add `Etiological:Genetic` because you know it is chromosomal; the name does not encode genetics.
- CORRECT: "Huntington's disease" → `Eponymous:Physician/Scientist` only. Do NOT add `Etiological:Genetic` even though it is hereditary.

**Taxonomy:**

1. **Eponymous:** Named after a person, real or fictional.
   * `Eponymous:Physician/Scientist`: named after the clinician or scientist who described it
   * `Eponymous:Other/Unknown`: named after a patient, mythological figure, literary character, or person of uncertain identity
2. **Geographical:** Named after a place (e.g., city, country, region). Use only when the name contains a place name.
3. **Anatomical:** The name specifies the primary body part affected.
   * `Anatomical:Organ`: a named organ (heart, liver, lung, kidney, etc.)
   * `Anatomical:System`: a body system (cardiovascular, respiratory, nervous, etc.)
   * `Anatomical:Tissue`: a tissue type (epithelial, connective, muscular, vascular, etc.)
   * `Anatomical:Cellular`: a cell type (lymphoblastic, erythrocytic, mast cell, etc.)

4.  **Etiological:** The name explicitly encodes a cause. Do not infer cause from background knowledge — only apply when a causal word or element is present in the name.
    *   `Etiological:Infectious Agent`: name contains a pathogen or infection term (bacterial, viral, fungal, parasitic)
    *   `Etiological:Substance-related`: name encodes a drug, chemical, toxin, or alcohol (e.g., "alcoholic", "drug-induced", "toxic")
    *   `Etiological:Injury-related`: name encodes trauma, radiation, or physical injury
    *   `Etiological:Genetic`: name explicitly encodes a hereditary or chromosomal cause or includes a specific gene name (e.g., "hereditary", "familial", "trisomy", "chromosomal", "FOXG1-related"). Do NOT apply to eponymous diseases that happen to be genetic.
    *   `Etiological:Nutritional`: name encodes a dietary deficiency or excess (e.g., "deficiency", "malnutrition", "vitamin")
    *   `Etiological:Immune-mediated`: name explicitly encodes an immune mechanism (e.g., "autoimmune", "immune-mediated", "hypersensitivity", "allergic")
    *   `Etiological:Occupational`: name encodes a work-related cause or exposure
    *   `Etiological:Procedural`: name encodes a medical or surgical procedure as cause (e.g., "post-operative", "iatrogenic", "post-procedural")
5.  **Pathophysiological:** The name encodes an abnormal biological process (e.g., infarction, fibrosis, necrosis, thrombosis, embolism, carcinoma, sclerosis). Distinct from Descriptive:Symptomatic — Pathophysiological is about the disease *process* itself, not a symptom experienced by the patient.
6.  **Descriptive:** The name describes a clinical feature, characteristic, or circumstance.
    *   `Descriptive:Symptomatic`: named for a symptom or clinical sign experienced by the patient (pain, fever, bleeding, cough, rash, paralysis). Distinct from Pathophysiological.
    *   `Descriptive:Metaphorical`: name uses a figurative or analogical term (e.g., "barrel chest", "strawberry tongue", "cri-du-chat")
    *   `Descriptive:Syndromic`: name explicitly uses the word "syndrome" or groups features under a collective label
    *   `Descriptive:Population`: name specifies the affected population (e.g., "juvenile", "neonatal", "maternal", "senile")
    *   `Descriptive:Course/Severity`: name encodes temporal course or severity (e.g., "acute", "chronic", "recurrent", "fulminant")
    *   `Descriptive:Congenital`: name explicitly uses "congenital" or equivalent terms encoding presence from birth
7.  **Socio-Behavioral:** Name encodes a social circumstance, psychological state, psychiatric condition, or behavioral pattern.
8.  **Vernacular/Obscure:** Name is a colloquial folk term, archaic clinical term, or foreign-language term not standard in modern medical English (e.g., "Caisson disease", "Shoshin", "Tabes", "Croup"). Do not apply this to widely-used modern medical terms just because they derive from Latin or Greek — e.g., 'Cholera', 'Pneumonia', 'Hepatitis' are standard terminology, not vernacular.
9.  **Other / Unclassified:** Name does not fit any other category.

**Input Format**

Entries follow the ICD-10 index format: "Main term > subterm" (e.g., "Alopecia > congenital"). Classify the full combined concept, not just the main term.

**Instructions:**

1. You will receive a numbered list of disease names.
2. For EACH name, assign one or more categories from the taxonomy above.
3. Use multiple labels only when different parts of the name independently encode different categories.
4. Return a single JSON object with the disease name as key and a list of category labels as value.
5. Use exact category label strings as shown in the taxonomy.

**Examples:**

**Input:**
1. "Addison's disease"
2. "Acculturation difficulty"
3. "Acute myocardial infarction"
4. "Iron deficiency anemia"
5. "Congenital alopecia"
6. "Autoimmune hepatitis"
7. "Huntington's disease"
8. "Hereditary spherocytosis"

**Output:**
{
"Addison's disease": ["Eponymous:Physician/Scientist"],
"Acculturation difficulty": ["Socio-Behavioral"],
"Acute myocardial infarction": ["Descriptive:Course/Severity", "Anatomical:Organ", "Pathophysiological"],
"Iron deficiency anemia": ["Etiological:Nutritional", "Descriptive:Symptomatic"],
"Congenital alopecia": ["Descriptive:Congenital", "Descriptive:Symptomatic", "Anatomical:Tissue"],
"Autoimmune hepatitis": ["Etiological:Immune-mediated", "Anatomical:Organ"],
"Huntington's disease": ["Eponymous:Physician/Scientist"],
"Hereditary spherocytosis": ["Etiological:Genetic", "Descriptive:Symptomatic"]
}